# Bandwidth scaling and spectral flatness enhancement of optical frequency combs from phase modulated continuous-wave lasers using cascaded four-wave mixing

V. R. Supradeepa,[1,2,*] and Andrew M. Weiner[1]

[1]*Electrical and Computer Engineering, Purdue University, 465 Northwestern Avenue, West Lafayette, IN 47907, USA*
[2]*Now at OFS Laboratories, 19 Schoolhouse Road, suite 105, Somerset, NJ 08873, USA*
*\*Corresponding author: supradeepa@ofsoptics.com*

We introduce a new cascaded four-wave mixing technique which scales up the bandwidth of frequency combs generated by phase modulation of a continuous wave laser while simultaneously enhancing the spectral flatness. As a result we demonstrate a 10 GHz frequency comb with over 100 lines in a 10-dB bandwidth in which a record 75 lines are within a flatness of 1-dB. The cascaded four-wave mixing process increases the bandwidth of the initial comb generated by modulation of a CW laser by a factor of five. The broadband comb has approximately quadratic spectral phase, which is compensated upon propagation in single mode fiber, resulting in a 10 GHz train of 940 fs pulses

Strong sinusoidal phase modulation of a continuous wave (CW) laser creates multiple sidebands leading to generation of a frequency comb [1]. Advantages of this technique are ability to create high repetition rate combs with stable optical center frequencies given by the source laser and convenient tuning of the repetition rate and optical center frequency. Such combs are a source of choice for applications in optical communications [2], radio frequency (RF) photonics [3, 4] and optical arbitrary waveform generation (OAWG) [5]. The bandwidth of such combs however is limited. The number of spectral lines scales linearly with the RF voltage driving the phase modulator. RF power handling limits the number of lines that can be generated by a single phase modulator, requiring a cascade of phase modulators to generate more lines. For example, state of the art commercially available, low Vpi phase modulators (Vpi ~3V) usually have a RF power limit of ~1W which limits the number of lines to ~20 in a 3-dB bandwidth (e.g., 200 GHz bandwidth at 10 GHz drive). To reach the 100 line level, we would have to cascade 5 modulators, which is both expensive and inefficient.

Furthermore, by phase modulation alone, the spectral flatness is quite poor, having significant line to line amplitude variations. A strongly modulated spectrum is undesirable for many applications. For example, for pulse train generation, line to line variations translate to reduced pulse quality. The flatness problem has been partially addressed by utilizing a Mach-Zehnder intensity modulator in series with the phase modulator [6]. An explanation for the improvement in spectral flatness provided by the addition of the intensity modulator was discussed in [7]. Figure 1 provides a schematic. The intensity modulator is driven such that it creates a flat-topped pulse (indicated by $a_1(t)$ in the figure). The action of the sinusoidal phase modulation in the window of the flat-topped pulse may be approximated to lowest order as a quadratic temporal phase. Strong quadratic phase in time performs a time to frequency mapping operation [8], creating a comb with spectral shape similar to the time domain pulse, which in this case is flat topped.

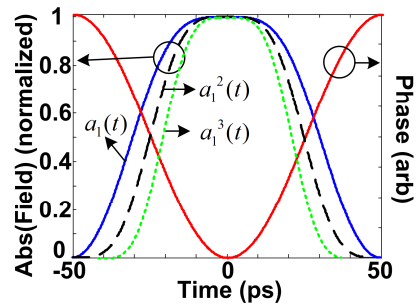

Fig. 1 Figure showing the time domain waveform and sinusoidal spectral phase in a cascaded intensity and phase modulator based comb generator.

However, owing to significant deviations of the sinusoid from a quadratic, the comb still has limited flatness with >5dB spectral variation between the lines in the central region. Recently, we proposed a technique to significantly flatten the spectrum: the duty factor of the flat-topped pulse was reduced utilizing two intensity modulators in series while the RF drive waveform to the phase modulator was shaped to better approximate a quadratic [9]. This allowed for significant improvement in spectral flatness to < 1dB. The bandwidth obtained however was limited by the number of phase modulators (~40 lines using 2 phase modulators). A method which scales bandwidth without increasing the number of modulators and improves flatness is desirable. There have been some methods to scale the bandwidth by first compressing the comb to a short pulse and then broadening the spectrum via nonlinear propagation in dispersion decreasing fiber or highly nonlinear fiber (HNLF) [10-12]. However, the spectral flatness of such combs is degraded. Also, owing to subtle interplay between dispersion and nonlinearity, the generated spectrum is not very stable.

In this work we introduce a simple scheme which can scale the bandwidth of the comb by several times (5, 7, -) in a stable and known fashion while simultaneously enhancing spectral flatness. By using just one intensity modulator and one phase modular, we generate a comb comprising over 100 lines within 10-dB bandwidth, out of which a record 75 lines are within the 1-dB bandwidth. Furthermore, our scheme allows compression of the comb into a bandwidth-limited train of 940 fs pulses via simple propagation in a dispersive fiber.

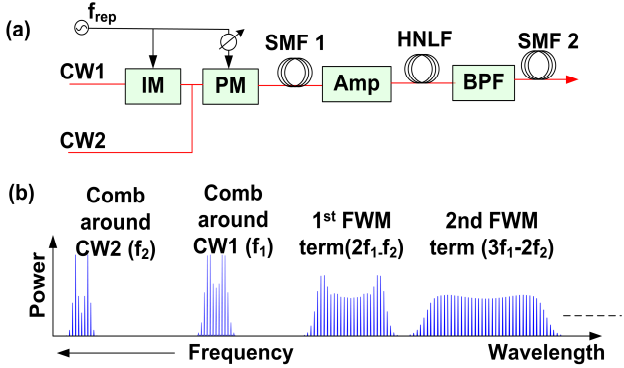

Fig. 2 (a) Experimental setup, CW – continuous wave laser, IM – Intensity modulator, PM – phase modulator, SMF – single mode fiber, HNLF – Highly nonlinear fiber, Amp – fiber amplifier, BPF – band pass filter, (b) Bandwidth scaling of the comb and enhanced spectral flattening

Fig. 2(a) shows the experimental setup. A CW laser (CW 1) at frequency $f_1$ is driven using an intensity and phase modulator to generate a partially flattened comb as discussed previously. If $a_1(t)$ (which is a flat-topped pulse) and $\phi(t)$ (which is a sinusoid) are the amplitude and phase modulation, the output after the IM and PM is $a_1(t)\exp(j\phi(t))$. We include a 2nd CW laser at frequency $f_2$ which is not amplitude modulated but which passes through the same phase modulator (its modulation is $\exp(j\phi(t))$). This is followed by a length of SMF whose length is chosen such that it delays the $f_2$ field by half an RF period (i.e., 50ps for a 10GHz drive frequency) relative to the $f_1$ field. For the $f_2$ field this transforms the phase modulation according to $\exp(j\phi(t)) \to \exp(-j\phi(t))$. The motivation is to ensure constructive bandwidth addition in the four wave mixing terms between the combs at $f_1$ and $f_2$. This is followed by a fiber amplifier followed by a near zero dispersion, low dispersion slope, highly nonlinear fiber (HNLF) and a band pass filter to select an appropriate frequency band. Assuming a short length of HNLF with near zero dispersion and low loss, the propagation regime is pure self-phase modulation which creates a cascade of four wave-mixing (FWM) terms. Looking towards the side of $f_1$, we will have new frequency components created at $2f_1 - f_2$, which would go as,

$$[a_1(t)\exp(j\phi(t))^2][\exp(-j\phi(t))^*] = a_1(t)^2 \exp(3j\phi(t)) \quad (1)$$

We clearly see that the bandwidth has tripled in this case. The next term in the cascade of four wave mixing terms will occur at $3f_1 - 2f_2$, which would be dominated by the term corresponding to mixing between the comb corresponding to first FWM term, the comb at $f_1$ and the comb at $f_2$, which goes as

$$[a_1(t)^2 \exp(3j\phi(t))][a_1(t)\exp(j\phi(t))][\exp(-j\phi(t))^*]$$
$$= a_1(t)^3 \exp(5j\phi(t)) \quad (2)$$

This indicates a bandwidth scaling of five times. Similarly, if we look at the higher order terms, we will have bandwidths scaling as 7 times, 9 times and so on. However, the efficiency reduces owing to increasing phase mismatch for the nonlinear process. An interesting aspect is that the amplitude coefficient of the above terms successively rises to higher powers of $a_1(t)$. Since $a_1(t)$ was chosen to be a flat-topped waveform, this raising to higher powers creates a reduction of the duty cycle of the time domain waveform by making the transition regions shorter, as depicted in Fig 1. By more effectively restricting the intensity to the region where the phase modulation is close to quadratic, more ideal time to frequency mapping is achieved, resulting in flatter combs. In the frequency domain, this can be viewed as successive convolution of the initial combs arranged such that the phases add constructively. This has the effect of smoothening out the spectra. Fig 2(b) shows schematically the bandwidth scaling and increasing spectral flatness in our cascaded four wave mixing scheme.

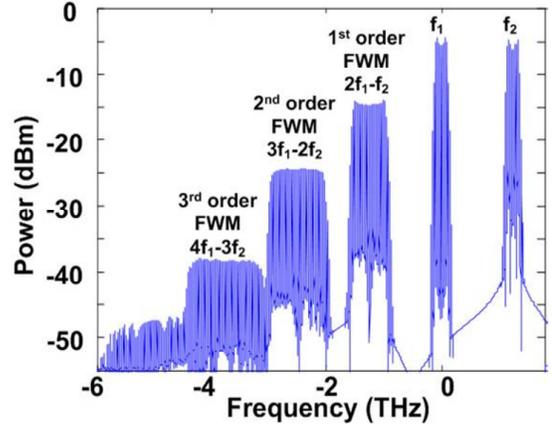

Fig. 3 Simulation showing bandwidth scaled flat comb generation

In our experiment, the two CW lasers were chosen to be at 1542nm ($f_1$) and 1532nm ($f_2$), respectively. The spacing was chosen such that there is no overlap between combs generated by adjacent FWM terms up to the 2nd order. The initial laser fields at 1532nm and 1542nm have an instantaneous linewidth of ~100 kHz and ~10 kHz respectively. The phase modulator has a Vpi of ~3V and is driven by an RF power of ~1W. The RF oscillator has a 10GHz frequency, and the first SMF spool is ~300m creating the required half-period (50ps) delay. We use a high power optical amplifier with ~1.5W output power. The HNLF we use has a length of 100m, D ~ 0 ps/nm/km and S = 0.02ps/nm^2/km. Practically, high conversion efficiency in FWM-based optical processing for lightwave communications has been demonstrated in several experiments using Watt level fiber amplifiers and commercial low dispersion

HNLF [for example 13-14]. Fig 3 shows the simulated output spectrum for our experiment obtained by a numerical simulation of the nonlinear Schrodinger equation taking into account 2$^{nd}$ and 3$^{rd}$ order dispersion in the fiber. The signal at 1542nm is taken to be at relative frequency of 0 THz. We can see a clear bandwidth scaling of successive terms of the FWM terms with progressive flattening of the power spectrum.

Fig 4(a) shows the measured comb around 1542nm with ~ 22 lines in a 10-dB bandwidth with mediocre flatness. The comb around 1532 nm is of similar bandwidth but is even less flat since it is generated by pure phase modulation (inset). Fig 4(b) shows the comb generated by the first FWM term centered around 1552nm. We can clearly see the 3 times bandwidth enhancement as well as the improvement in spectral flatness compared to the initial comb. Fig 4(c) and 4(d) shows the 2nd FWM term centered on 1562 nm in linear and log scale. We get around 20 mW of power in this spectral region. We clearly see further bandwidth enhancement to 5 times and the significant improvement in spectral flatness. The new comb has >100 lines in a 10-dB bandwidth with a record 75 of them within 1-dB. We have thus generated flat combs in a convenient fashion without the need for temporal shaping of the phase modulator waveforms as done in [9].

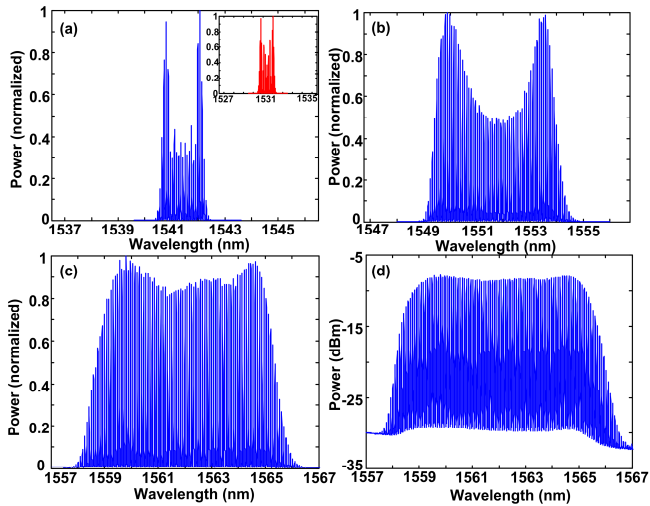

Fig. 4. Experimental results - (a) Input comb (purely phase modulated 2nd comb shown in the inset), (b) comb generated by the 1st order FWM term, (c) comb generated by the 2nd order FWM term, (d) 2nd order comb in log scale.

The combs generated in this way have temporal phase very close to quadratic. By Fourier transform relations, the transform of a purely quadratic temporal phase waveform is a quadratic spectral phase. This means that the generated comb has a linear chirp which can easily compensated using dispersive media, e.g., a length of single mode fiber. This allows us to create high quality short pulses from the generated combs with a simple compression scheme. Fig 5(a) shows the spectral phase of the comb generated at the 2$^{nd}$ order FWM term measured via the half-repetition rate modulation scheme described in [15]. We see that the measured spectral phase fits very well to a quadratic. The pulse is compressed using ~200m of SMF. The spectral phase is again measured via the scheme of [15] and used to compute the compressed pulse temporal intensity profile, Fig 5(b). The compressed pulse has a FWHM duration of ~940fs and is in close agreement with the time domain intensity simulated using the measured power spectrum and assuming flat spectral phase.

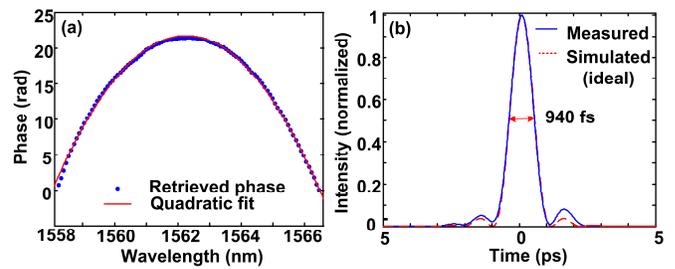

Fig. 5. (a) Measured spectral phase of the comb from the 2$^{nd}$ order FWM term and quadratic fit, (b) Measured time domain intensity of the comb after phase correction and simulated ideal time domain intensity assuming a flat spectral phase for the measured spectrum.

In summary, we have demonstrated a simple scheme to significantly scale the bandwidth of phase modulated CW combs while enhancing spectral flatness. Our experiment yielded a 10GHz comb with >1THz of bandwidth (with >750 GHz in a 1-dB bandwidth) using just a single intensity and phase modulator. Our scheme preserves electro-optic comb generator advantages such as tunability of the optical center frequency and the repetition rate. Furthermore, our scheme simply supports high quality pulse compression, resulting in the current work in trains of bandwidth-limited 940 fs pulses.

This work was supported in part by the National Science Foundation under grant ECCS-1102110 and by the Naval Postgraduate School under grant N00244-09-1-0068 under the National Security Science and Engineering Faculty Fellowship program.

**References**
1. H. Murata, A. Morimoto, T. Kobayashi, and S. Yamamoto, IEEE J. Sel. Top. Quantum Electron. 6, 1325–1331 (2000).
2. Takuya Ohara, Hidehiko Takara, Takashi Yamamoto, Hiroji Masuda, Toshio Morioka, Makoto Abe, and Hiroshi Takahashi, J. Lightwave Technol. 24, 2311- (2006).
3. J. Capmany and D. Novak, Nat. Photonics 1, 319-330 (2007).
4. V. R. Supradeepa, Christopher. M. Long, Rui Wu, Fahmida Ferdous, Ehsan Hamidi, Daniel E. Leaird and Andrew M. Weiner, Nature Photon. 6, 186-194 (2012)
5. Z. Jiang, C. B. Huang, D. E. Leaird, and A. M. Weiner, Nature Photon. 1, 463-467 (2007).
6. Masamichi Fujiwara, Mitsuhiro Teshima, Jun-ichi Kani, Hiro Suzuki, Noboru Takachio, and Katsumi Iwatsuki, J. Lightwave Technol. 21, 2705- (2003).
7. Víctor Torres-Company, Jesús Lancis, and Pedro Andrés, Opt. Lett. 33, 1822-1824 (2008).
8. Brian H. Kolner and Moshe Nazarathy, Opt. Lett. 14, 630-632 (1989).
9. Rui Wu, V. R. Supradeepa, Christopher M. Long, Daniel E. Leaird, and Andrew M. Weiner, Opt. Lett. 35, 3234-3236 (2010).
10. C. B. Huang, S. G. Park, D. E. Leaird, and A. M. Weiner, Optics Express, vol. 16, 2520-2527, 2008.
11. R. P. Scott, N. K. Fontaine, J. P. Heritage, B. H. Kolner, and S. J. B. Yoo, OFC 2007.
12. K. R. Tamura, H. Kubota, and M. Nakazawa, IEEE J. Quantum Electron. , vol. 36, 773-779, 2000.
13. C. S. Bres, N. Alic, E. Myslevits and S. Radic, J. Lightwave Tech 27, 356-363, 2009.
14. P. Devgan, R. Tang, V. S. Grigoryan and P. Kumar, J. Lightwave Tech 24, 3677-3683, 2006.
15. V. R. Supradeepa, C. M. Long, D. E. Leaird, and A. M. Weiner, Optics Express, vol. 18, 18171-18179, 2010.